\documentclass[onecolumn,secnumarabic,amssymb, nobibnotes]{revtex4-2}
\usepackage{times}
\usepackage[pdftex]{graphicx}
\usepackage{subfig}
\usepackage{bm}
\usepackage{amsmath}
\usepackage{indentfirst}
\usepackage{float}
\usepackage[colorlinks]{hyperref}
\usepackage[no-test-for-array]{nicematrix}
\usepackage{tikz}
\usepackage{lipsum}
\usepackage{makecell}
\usepackage[normalem]{ulem}
\newcommand{\Hop}{\hat{H}}

\newcommand{\aops}{\hat{a}}
\newcommand{\adop}{\hat{a}^{\dagger}}
\newcommand{\paras}{\vec{\alpha}}

\newcommand{\loss}{{\rm loss}}
\newcommand{\norm}{{\rm norm}}

\newcommand{\im}{{\rm i}}

\begin{document}

\title{Phonon Number Measurement Using Optimal Composite Pulses}




\author{Xie-Qian Li}
\affiliation{Institute for Quantum Science and Technology, National University of Defense Technology, Changsha Hunan 410073, China}
\affiliation{Hunan Key Laboratory of Mechanism and Technology of Quantum Information, Changsha Hunan 410073, China}

\author{Chun-Wang Wu}
\email{cwwu@nudt.edu.cn}
\affiliation{Institute for Quantum Science and Technology, National University of Defense Technology, Changsha Hunan 410073, China}
\affiliation{Hunan Key Laboratory of Mechanism and Technology of Quantum Information, Changsha Hunan 410073, China}

\author{Ping-Xing Chen}
\email{pxchen@nudt.edu.cn}
\affiliation{Institute for Quantum Science and Technology, National University of Defense Technology, Changsha Hunan 410073, China}
\affiliation{Hunan Key Laboratory of Mechanism and Technology of Quantum Information, Changsha Hunan 410073, China}

\begin{abstract}
{The measurement of phonon numbers in laser-cooled ions is pivotal for determining whether an ion is in its ground state. Commonly employed experimental techniques include the analysis of red-to-blue sideband ratios and the adiabatic evolution on the red-sideband transition. We introduce a theoretical approach utilizing composite pulses that enables the direct measurement of the population in a specific Fock state, eliminating the need for state evolution fitting. This method offers a more straightforward quantification of higher Fock state populations compared to the red-sideband adiabatic evolution technique. By applying quantum optimal control strategies, we enhance the fidelity of the unitary operations executed by the composite pulses. \textcolor{black}{By utilizing numerical calculation methods of quantum optimal control, we have overcome the limitations of analytical calculations that require weak coupling and first-order Lamb-Dicke expansion, allowing our method to achieve sufficient accuracy in strong coupling regime.} Furthermore, we also present \textcolor{black}{a data processing method that allows the estimated value of the phonon number to be closer to the true value.} An illustrative example of its efficacy in high Fock state measurement is provided.}
\end{abstract}

\date{June 25, 2024}
\maketitle

\vspace{8mm}

\section{Introduction}\label{sec:introduction}
Trapped-ion system is one of the most promising candidates for quantum computing \cite{Wineland:1998uh, Kielpinski:2002aa, Cirac:2000aa, Chiaverini:2004aa}, and there has been rapid progress in this direction recently \cite{Pino:2021tf, Elben:2020aa, Marciniak:2022aa, Brydges:2019aa, Joshi:2022aa, McCormick:2019aa, Todaro:2021aa, Zhang:2019aa, Lu:2019aa, Wang:2021aa}. To achieve high-quality qubits, it is crucial to initially cool the trapped ions. Consequently, evaluating the cooling performance of a scheme necessitates the accurate measurement of the resultant phonon number.

In an ion trap system, the measurement of external states (phonon states) is achieved by transferring their information into internal states (qubits). \textcolor{black}{Sideband-ratio thermometry  is a commonly used method} \cite{An:2015aa, PhysRevResearch.5.023022, Poschinger2011QuantumOE, RevModPhys.75.281}, \textcolor{black}{which assumes the phonon distribution is the thermal state. Thus,} this method only provides the average phonon number $\bar{n}$ and does not reveal the occupation number of each Fock state. Another method to obtain the occupation number of each Fock state sequentially is by using adiabatic evolution on the red-sideband transition to excite $\vert g\rangle \vert n \rangle$ to $\vert e\rangle \vert n-1 \rangle$ (where $\vert g\rangle$ and $\vert e\rangle$ denote the ground and excited states of the qubit and $\vert n \rangle$ denotes the Fock state), followed by the measurement of the occupation probability of $\vert g\rangle \vert 0 \rangle$. If the population is detected in $\vert g\rangle \vert 0 \rangle$, the measurement cycle stops and the occupation of the phonon can be inferred as $|0\rangle$. If no occupation is detected, the phonon occupies nonzero Fock numbers. Then a carrier laser pulse pumps the state $\vert e\rangle \vert n-1 \rangle$ to $\vert g\rangle \vert n-1 \rangle$, and the next measurement cycle begins. By this protocol, the occupation numbers of higher Fock states can be obtained by repeating these carrier pulses, adiabatic red-sideband pulses and detection processes \cite{Wolf:2019aa, An:2015ab, Um:2016aa}.

 Composite pulse technology, initially employed in nuclear magnetic resonance (NMR) as detailed by Levitt (1986)\cite{LEVITT198661}, has been further developed through the application of quantum optimal control techniques\cite{Hansen2007, Tosner2009}. More recently, the use of composite pulses has been extended into the realm of quantum coherent manipulations, significantly enhancing the gate fidelity and robustness\cite{PhysRevA.93.023830, PhysRevA.84.065404, PhysRevA.107.023103, PhysRevLett.118.133202}. Additionally, the protocol described by Um et al. (2016)\cite{Um:2016aa} was adapted by employing composite pulses in place of adiabatic red-sideband evolution, as demonstrated in a subsequent study\cite{PhysRevA.100.060301}. \textcolor{black}{Composite pulses achieve the suppression of unwanted transitions and the enhancement of desired transitions by adjusting multiple parameters of the pulse, such as detuning, duration, and initial phase, so that these parameters are matched. In other words, composite pulses can narrow the coupling region of Fock state transitions, focusing the transitions between the energy levels we desire. Theoretically, more pulses can more easily be designed to closely meet the requirements of the composite pulse.} \textcolor{black}{Mallweger et al. (2024)\cite{Mallweger2024} performed experiments of phonon measurements using ultra-narrowband composite pulses. The focus of our work emphasizes the specific application of numerical optimization methods in composite pulse design. The transitions excited by composite pulses are not limited to first-order sidebands, and we propose correction schemes for situations where the pulses are imperfect.}

 We propose a theoretical method to measure phonon numbers using composite pulses. This approach allows for the selective excitation of specific states to an excited state while maintaining other Fock states in the ground state. Consequently, it enables the correlation between external and internal states, achieving the direct measurement of phonon numbers. Unlike the method that relies on red-to-blue sideband ratios, our approach does not require the measurement and fitting of evolution curves. Additionally, compared to schemes that use adiabatic evolution, our method can directly measure high Fock state populations, such as directly assessing the population of $\vert g \rangle \vert 20 \rangle $ without measuring intermediate states from $\vert g \rangle \vert 0 \rangle $ to $\vert g \rangle \vert 19 \rangle$ \textcolor{black}{by numerically designing the composite pulses which only excite $\vert g \rangle \vert 20 \rangle $ to $\vert e \rangle$.} Our technique avoids the sequential use of red-sideband and carrier pulses, relying instead on the phase modulation of a single laser beam, which may offer higher fidelity in measuring high Fock states. We employ \textcolor{black}{quantum optimal control technique} \cite{Werschnik2007QuantumOC} to enhance the manipulation fidelity of composite pulse operations. Specifically, we use a configuration similar to traditional composite pulse techniques but apply optimal quantum control to determine the best parameters for the pulses, including coupling strength, phase, and detuning. Our method is not limited by first-order approximations or weak coupling conditions. Furthermore, we introduce a strategy to mitigate the effects of dimension truncation and fidelity issues, thereby enhancing the measurement accuracy.

This paper is organized as follows: Sec.~\ref{sec:method} outlines our approach for optimizing the parameters of composite pulses through quantum optimal control. Sec.~\ref{sec:SWAP} demonstrates the application of our method by optimizing the SWAP operation for phonon number measurement. In Sec.~\ref{sec:strong_cp}, we focus on optimizing composite pulses within the strong coupling regime. Sec.~\ref{sec:modify} details modifications of our approach to enhance precision, providing examples in both weak and strong coupling regimes, and discusses its application in high Fock state measurement. We conclude in Sec.~\ref{sec:summary}.

\section{Scheme of quantum optimal control composite pulses}\label{sec:method}
We consider a model where an ion trapped in a harmonic potential interacts with a laser. When the laser is treated as a classical field, the model can be equivalently described as an interaction between the ion's internal state (a two-level system) and its external state (phonons). The dynamics of this process are governed by the Schrödinger equation, with the Hamiltonian
\begin{align}\label{eq:h}
\Hop(\Delta, \Omega, \phi) = -\Delta \vert e\rangle\langle e\vert + \nu \adop\hat{a} + \frac{\Omega}{2}\left(e^{\im\phi} \vert e\rangle\langle g\vert e^{-\im \eta(\adop + \hat{a})} +e^{-\im\phi} \vert g\rangle\langle e\vert e^{\im \eta(\adop + \hat{a})}\right) ,
\end{align}
where the Lamb-Dicke parameter is defined as \textcolor{black}{$\eta=k\sqrt{\hbar/(2m\nu)}$}, $\nu$ is the \textcolor{black}{trap frequency}, $k$ is the wavevector of the laser, $m$ is the ion's mass, $\Delta$ is the laser detuning, $\Omega$ is the coupling strength, $\phi$ is the laser initial phase, $\adop$ and $\aops$  are creation and annihilation operators of phonons. We consider a situation where we use $^{40}Ca^+$ as the trapped ion, the trap frequency $\nu=2\pi*1$MHz is nondimensionalized to 1, and other parameters are scaled with it. In the numerical computation, we set  $\hbar=1$, $\eta=0.084$ and $\nu=1$.

\textcolor{black}{A single laser pulse interacts with an ion under a set of parameters including detuning $\Delta$, coupling strength $\Omega$ and initial phase $\phi$ for a certain duration $t$. Conventional single laser pulses, e.g. the carrier transition pulse or the first-order sideband transition pulse, will excite the internal state of the ion with different Fock states simultaneously. However, for phonon states distributed in a certain way, we often need a method to directly and quickly read the population of specific Fock states. In this case, we need to design an evolution that selectively pumps the Fock state of interest, which has its internal state in the ground state, completely to the excited state, while leaving other Fock states completely unpumped, thus converting the population information of the external state into the internal state for direct measurement. A single laser pulse clearly cannot achieve this. However, by adjusting the parameters of the Hamiltonian in Eq.~\ref{eq:h}, such as their duration, detuning, and relative phases, we can construct a composite pulse that narrows the regime of Fock-state-dependent coupling strength and suppress the unwanted transitions.}

\begin{figure}
 \centering
  \includegraphics[width=1.0\columnwidth]{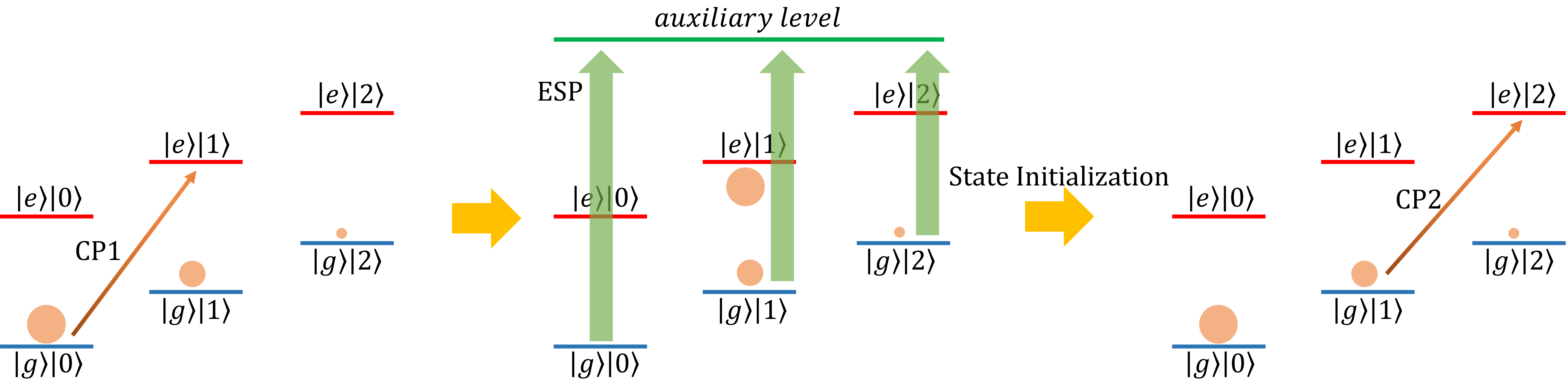}
  \caption{Steps to measure Fock state population using composite pulses (CP). \textcolor{black}{The orange arrow represents the composite pulse laser; the circles above the ground state energy level represent the population of phonon Fock states, for instance, the population of phonons in a thermal state rapidly decreases with increasing Fock energy levels; the green arrow represents the state detection laser.} We can measure the populations of $\vert g\rangle \vert 0 \rangle $, $\vert g\rangle \vert 1 \rangle $, and $\vert g\rangle \vert 2 \rangle $ either sequentially or non-sequentially. For example, to measure the population of $\vert g\rangle \vert 0 \rangle $, we first employ composite pulses to transit $\vert g\rangle \vert 0 \rangle $ to $\vert e\rangle \vert 1 \rangle $. Subsequently, we utilize electron shelving pulse (ESP) to determine the population of $\vert g\rangle \vert 0 \rangle $. The state is then reinitialized before applying the next composite pulse.}
 \label{fig:fig1}
\end{figure}

\textcolor{black}{For example,} we employ three pulses to manipulate the phonon, which collectively act as a unitary operator,
  \begin{align}\label{eq:evolution}
U_{p3}U_{p2}U_{p1}=e^{-\im\Hop(\paras_3)t_3}e^{-\im\Hop(\paras_2)t_2}e^{-\im\Hop(\paras_1)t_1}
 =e^{-\im\Hop(\Delta_3, \Omega_3, \phi_3)t_3}e^{-\im\Hop(\Delta_2, \Omega_2, \phi_2)t_2}e^{-\im\Hop(\Delta_1, \Omega_1, \phi_1)t_1}.
 \end{align}
 These composite pulses are designed to measure the phonon number. Initially, the ion is cooled to its ground state with no excited population. The application of composite pulses evolves the state from $\vert g\rangle \vert 0 \rangle$ to $\vert e\rangle \vert 1 \rangle$, while leaving $\vert g\rangle \vert n \rangle (n\ne 0)$ unaffected. Following this, the state $\vert g\rangle \vert 0 \rangle$ is excited, and its population is determined using the electronic shelving method \cite{2000Controlling}. The population of $\vert g\rangle \vert 1 \rangle$ is similarly assessed by exciting this state and keeping the other $\vert g\rangle \vert n \rangle$ states unchanged. This procedure is iterated to ascertain the populations of various Fock states. Figure~\ref{fig:fig1} provides a schematic diagram that illustrates the truncation of the Fock state space to three levels. Consequently, the primary challenge is to engineer composite pulses that selectively excite the intended Fock state without influencing other states.  \textcolor{black}{Typically, this composite pulse consists of multiple blue-sideband pulses. The coupling strength of different sideband processes varies, and by utilizing the phase differences between multiple blue-sideband pulses, we can make the probability of unwanted transitions zero.}
 
We employ quantum optimal control to design three pulses that excite selected Fock states to an excited state. The parametric Hamiltonian is denoted as $\Hop(\paras)$, where $\paras$ represents a list of tunable parameters. Numerical optimization techniques are utilized to minimize the loss function as defined in Eq. (\ref{eq:loss}),
 
 \begin{align}\label{eq:loss}
\loss(\paras_1, \dots, \paras_n,\vec{t}) = \norm( e^{-\im\Hop(\paras_n)t_n} \dots e^{-\im\Hop(\paras_1)t_1}-\textcolor{black}{U_{target}})= \norm( U_{pn} \dots U_{p1}-\phi_G \textcolor{black}{U'_{target}}),
\end{align}
where \textcolor{black}{$U_{target}$} is the target manipulation, $\phi_G$ is the global phase of the unitary operation, and \textcolor{black}{$U'_{target}$} is the target manipulation with arbitrary phase.

 The loss function measures the Euclidean distance (norm) between the unitary matrix of the composite pulse and the target \textcolor{black}{transition}. It is important to note that the global phase does not alter the functionality of the \textcolor{black}{transition}. Consequently, in subsequent computations, we employ Eq.  (\ref{eq:loss2}), which converts each complex element in the matrix to its magnitude,

\begin{align}\label{eq:loss2}
\loss(\paras_1, \dots, \paras_n,\vec{t}) = &\norm (\norm.( e^{-\im\Hop(\paras_n)t_n} \dots e^{-\im\Hop(\paras_1)t_1})-\norm.(\textcolor{black}{U_{target}})),
\end{align}
where $\rm{norm.}(\cdot)$ transfers every element of the matrix into its \textcolor{black}{modulus}.     

In this study, the loss function is defined by specifying the target \textcolor{black}{transition} and the initial parameter values. By numerically minimizing the loss function, we can determine the optimal parameter values. The optimization techniques employed include the limited-memory Broyden-Fletcher-Goldfarb-Shanno (L-BFGS) algorithm \cite{Liu:1989aa} and the particle swarm optimization algorithm \cite{Kennedy:2010aa}. All programming was performed using Julia \cite{julia}.

\section{Phonon number measurement using optimized SWAP operation}\label{sec:SWAP}

Initially, we propose a scheme utilizing a series of SWAP operations \cite{Gulde:2003aa, Schmidt-Kaler:2003aa} to implement the process depicted in Fig.~\ref{fig:fig1} using quantum optimal control in the weak coupling regime. 

In Ref.~\cite{Gulde:2003aa, Schmidt-Kaler:2003aa}, composite pulses were used to implement the SWAP operation, considering only a three-dimensional Boson space as described in \textcolor{black}{table}~(\ref{eq:SWAP}),
\begin{align}\label{eq:SWAP}
  \setlength{\arraycolsep}{0.5pt}
  \begin{bNiceMatrix}[first-row, first-col]
  \CodeBefore
  \cellcolor{red!15}{1-1,1-5,5-1,5-5}
  \Body
    & \langle g\vert  \langle 0 \vert & \langle g\vert  \langle 1 \vert & \langle g\vert  \langle 2 \vert & \langle e\vert  \langle 0 \vert & \langle e\vert  \langle 1 \vert & \langle e\vert  \langle 2 \vert\\
  \vert g\rangle \vert 0\rangle & 0 & 0 & 0 & 0 & 1 & 0 \\
  \vert g\rangle \vert 1\rangle & 0 & 1 & 0 & 0 & 0 & 0 \\
  \vert g\rangle \vert 2\rangle & 0 & 0 & 1 & 0 & 0 & 0 \\
  \vert e\rangle \vert 0\rangle & 0 & 0 & 0 & 1 & 0 & 0 \\
  \vert e\rangle \vert 1\rangle & 1 & 0 & 0 & 0 & 0 & 0 \\
  \vert e\rangle \vert 2\rangle & 0 & 0 & 0 & 0 & 0 & 1
  \end{bNiceMatrix},
\end{align}
and their theoretical analysis only considered a perfect blue-sideband situation like Eq.(\ref{eq:perfect_bsb}),
\begin{align}\label{eq:perfect_bsb}
R_{BSB}(\theta,\phi)={\rm exp}[\im \frac{\theta}{2}(e^{\im \phi}\hat{\sigma}^+\adop +e^{-\im \phi}\hat{\sigma}^-\aops)].
\end{align}
 \textcolor{black}{Eq.~\ref{eq:perfect_bsb} only considers the situation where the blue-sideband transition can be perfectly executed. If we substitute the analytical parameters derived from Eq.~\ref{eq:perfect_bsb} into our numerical expression, which has not undergone the Lamb-Dicke first-order approximation and the weak coupling approximation, we can observe that such composite pulses have a certain deviation from the ideal transition target.} Specifically, the norm of the unitary matrix elements [${\rm norm}(U)$] corresponding to the 3-pulse composite pulses in \textcolor{black}{table}~(\ref{eq:SWAP_cj}) will be affected \textcolor{black}{(due to space constraints, we illustrate our scheme using a four-dimensional Boson space cutoff)},
\begin{align}\label{eq:SWAP_cj}
\arrayrulecolor{cyan}
\begin{bNiceArray}{ccc|ccc}[margin]
\CodeBefore
\cellcolor{red!15}{1-5,5-1}
\Body
0.587 	&	0.060 	&	0.005 	&	0.074 	&	0.803 	&	0.037 	\\
0.027 	&	0.906 	&	0.022 	&	0.003 	&	0.074 	&	0.416 	\\
0.002 	&	0.045 	&	0.997 	&	0.000 	&	0.002 	&	0.059 	\\
\hline
0.080 	&	0.005 	&	0.000 	&	0.996 	&	0.050 	&	0.003 	\\
0.805 	&	0.052 	&	0.003 	&	0.058 	&	0.588 	&	0.021 	\\
0.021 	&	0.415 	&	0.071 	&	0.000 	&	0.037 	&	0.906 	
\end{bNiceArray},
\end{align}
where in the weak coupling regime, we set
\begin{align}
&\Omega_1=\Omega_2=\Omega_3=0.1, \nonumber\\
&\Delta_1=\Delta_2=\Delta_3=1,\ \phi_1=0\nonumber.
 \end{align}
\textcolor{black}{Here, we set $\Delta$ as the constant 1 and $\Omega$ as 0.1, which do not participate in the optimization. This confines the transition to a blue sideband transition. The optimal parameters can be derived using analytical methods:}
\begin{align}
 &t_1=\frac{\pi}{\sqrt{2}\eta\Omega_1}=264.46,\ t_2=\frac{\sqrt{2} \pi}{\eta\Omega_2}=528.91,\ t_3=\frac{\pi}{\sqrt{2}\eta\Omega_3}=264.46,\nonumber\\
 &\phi_2={\rm arccos}[{\rm cot^2}(\pi/\sqrt 2)]=0.95, \ \phi_3=0.\nonumber
\end{align}

\textcolor{black}{However, the analytical method is based on the first-order approximation of Lamb-Dicke and ignores the issues of AC Stark shifts caused by laser coupling. In numerical computations, we employ Eq. (\ref{eq:h}) and Eq. (\ref{eq:evolution}) to represent the evolution of composite pulses, and use Eq.~(\ref{eq:SWAP}) as $U_{target}$.} Subsequently, we utilize the loss function Eq. (\ref{eq:loss2}) for numerical optimization. The modulus of the unitary matrix elements, denoted as ${\rm norm.}(U)$, for the optimized SWAP operation is given by \textcolor{black}{table} (\ref{eq:SWAP_op}),

\begin{align}\label{eq:SWAP_op}
\arrayrulecolor{cyan}
\begin{bNiceArray}{ccc|ccc}[margin]
\CodeBefore
\cellcolor{red!15}{1-5,5-1}
\Body
0.011 	&	0.049 	&	0.005 	&	0.031 	&	0.998 	&	0.034 	\\
0.042 	&	0.997 	&	0.047 	&	0.004 	&	0.050 	&	0.017 	\\
0.005 	&	0.047 	&	0.999 	&	0.000 	&	0.003 	&	0.008 	\\
\hline
0.011 	&	0.003 	&	0.000 	&	0.999 	&	0.031 	&	0.003 	\\
0.998 	&	0.042 	&	0.006 	&	0.012 	&	0.009 	&	0.043 	\\
0.042 	&	0.017 	&	0.008 	&	0.002 	&	0.035 	&	0.998 		
\end{bNiceArray},
\end{align}
and the optimized parameters of the composite pulses are
\begin{align}
t_1&=	267.38 	, \ 
t_2=	529.39 	,\ 
t_3=	263.81 	, \nonumber\\
\Delta_1&=	1.00 	,\ 
\phi_2=	1.10 	, \
\phi_3=	0.41 	. \nonumber
\end{align}
The fidelity of the SWAP operation is notably higher, with the laser phases $\phi_2$ and $\phi_3$ playing crucial roles in determining the fidelity.

\begin{figure}[htbp]
 \centering
 \subfloat[]
  {
      \label{fig:subfig_t}\includegraphics[width=0.5\textwidth]{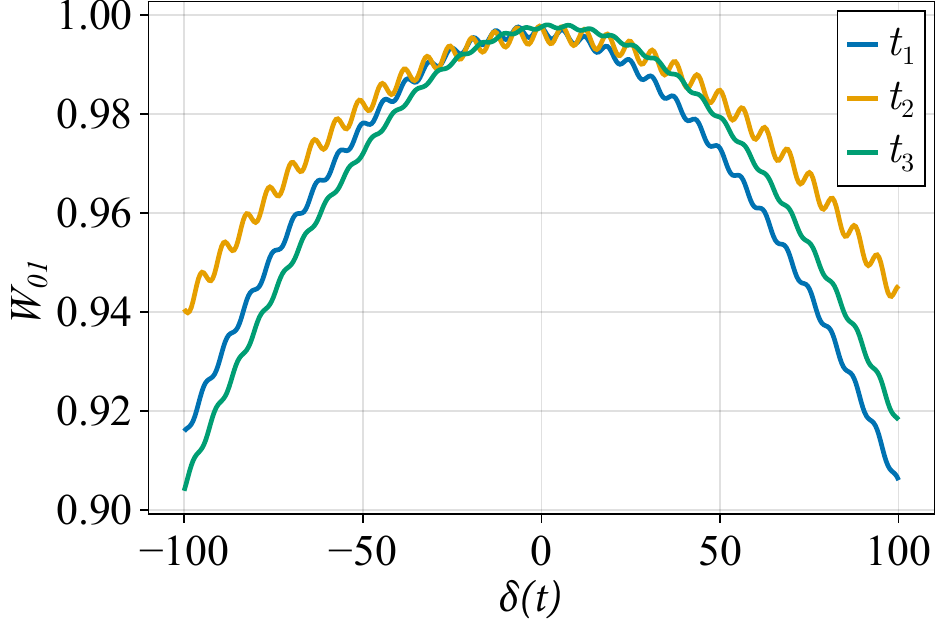}
  }
\subfloat[]
  { 
      \label{fig:subfig_phi}\includegraphics[width=0.5\textwidth]{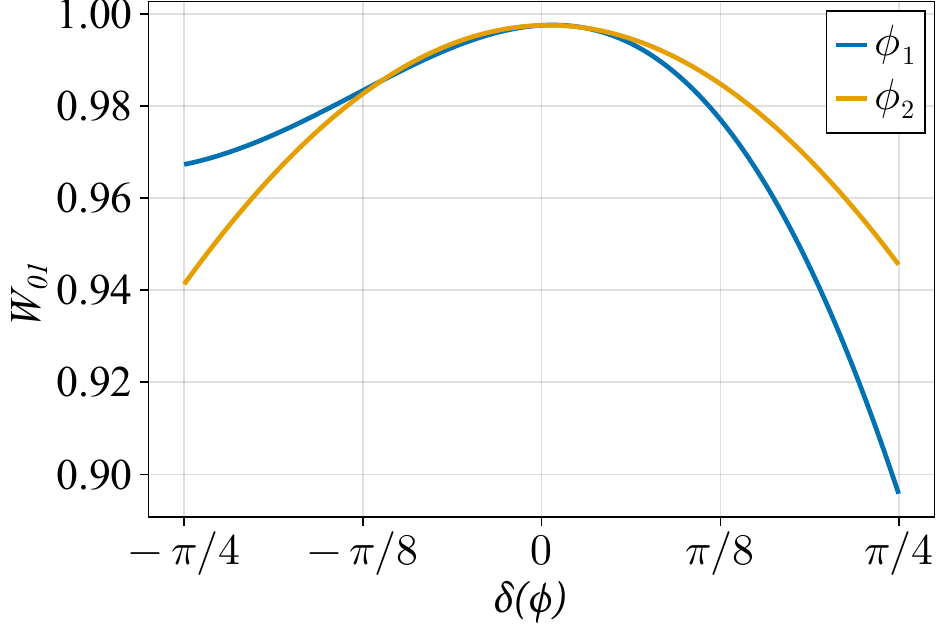}
  }
  \caption{The \textcolor{black}{transition probability of Fock state $|0\rangle$ to $|1\rangle$} varies with time detuning (a) and phase detuning (b). \textcolor{black}{Near the optimal values of the parameters, we can allow for temporal and phase errors in the experiment, which can easily achieve 99\% transition probability with the experimental instruments. The wave-like pattern observed in (a) is due to phase errors caused by time deviations when the laser is turned off.}}   
  \label{fig:fig22} 
 \end{figure}

We provide a brief overview of the robustness of composite pulses. Figures~\ref{fig:subfig_t} and \ref{fig:subfig_phi} show \textcolor{black}{that the transition probability of Fock state $|0\rangle$ to $|1\rangle$ ($W_{01}$) changes with the duration deviation ($\delta(t)$) and phase deviation ($\delta(\phi)$).} \textcolor{black}{The transition probability still exceeds 99\%} when $|\Delta t| <10 \mu s$ and $|\Delta\phi| <\frac{\pi}{4}$. Achieving this level of accuracy is feasible experimentally using an arbitrary waveform generator. Therefore, the robustness of the composite pulse is sufficient for thermometry applications.
 
We utilized the quantum optimal control method to design phonon measurement pulses. Primarily, the phonon state after cooling is the Fock state $\vert 0 \rangle$. Assuming the cooled ion only occupies populations near the Fock state $\vert 0 \rangle$, we truncated the Bosonic space dimensions to four. Consequently, we developed three composite pulses to facilitate the measurement of phonon Fock states. These pulses individually induce transitions from $\vert g\rangle \vert 0 \rangle$ to $\vert e\rangle \vert 1 \rangle$, $\vert g\rangle \vert 1 \rangle$ to $\vert e\rangle \vert 2 \rangle$, and $\vert g\rangle \vert 2 \rangle$ to $\vert e\rangle \vert 3 \rangle$.

We configured the coupling strength in a weak coupling regime, setting $\Omega_1$, $\Omega_2$, and $\Omega_3$ as 0.1. Under these conditions, the AC Stark shift, as reported by \cite{PhysRevA.49.2771}, is minimal; thus, we set $\Delta_1$, $\Delta_2$, and $\Delta_3$ as 1. The phases $\phi_1$, $\phi_2$, and $\phi_3$ were considered, with $\phi_1$ consistently as zero, and $\phi_2$ and $\phi_3$ as relative phases to $\phi_1$. We then optimized $t_1$, $t_2$, $t_3$, $\phi_2$, and $\phi_3$ using a particle swarm optimization algorithm to minimize the loss function and avoid local minima. Subsequently, the L-BFGS algorithm was employed to refine these parameters to the optimal values of the loss function. We confined the parameter search within the ranges $t \in [0,\frac{4\pi}{\eta\Omega}]$ and $\phi \in [0,2\pi]$.

We design composite pulses using quantum optimal control techniques. The phase of the unitary transformation in composite pulses does not affect the population transfer between states $\vert g \rangle \vert n\rangle$, hence we present only the norm of $U_{CP}$. After the optimization, the ${\rm norm.}(U_{CP})$ of the composite pulse pumping $\vert g \rangle \vert 0\rangle$ to $\vert e \rangle \vert 1\rangle$ is
\begin{align}
\arrayrulecolor{cyan}
\begin{bNiceArray}{cccc|cccc}[margin]
\CodeBefore
\cellcolor{red!15}{1-6,6-1}
\Body
0.060 	&	0.000 	&	0.000 	&	0.000 	&	0.000 	&	0.998 	&	0.000 	&	0.000 	\\
0.000 	&	0.999 	&	0.000 	&	0.000 	&	0.000 	&	0.000 	&	0.037 	&	0.000 	\\
0.000 	&	0.000 	&	0.999 	&	0.000 	&	0.000 	&	0.000 	&	0.000 	&	0.053 	\\
0.000 	&	0.000 	&	0.000 	&	1.000 	&	0.000 	&	0.000 	&	0.000 	&	0.000 	\\
\hline
0.000 	&	0.000 	&	0.000 	&	0.000 	&	1.000 	&	0.000 	&	0.000 	&	0.000 	\\
0.998 	&	0.000 	&	0.000 	&	0.000 	&	0.000 	&	0.060 	&	0.000 	&	0.000 	\\
0.000 	&	0.037 	&	0.000 	&	0.000 	&	0.000 	&	0.000 	&	0.999 	&	0.000 	\\
0.000 	&	0.000 	&	0.053 	&	0.000 	&	0.000 	&	0.000 	&	0.000 	&	0.999 		
\end{bNiceArray},\nonumber
\end{align}
and the optimal parameters are 
\begin{align}
t_1&=470.00, \ t_2=1262.16, \ t_3=1247.93;\nonumber\\
\phi_2&=4.53, \ \phi_3=6.77. \nonumber
\end{align}
The normalized $U_{CP}$ of the composite pulse, which transitions from $|g\rangle |1\rangle$ to $|e\rangle |2\rangle$ after optimization, is
\begin{align}
\arrayrulecolor{cyan}
\begin{bNiceArray}{cccc|cccc}[margin]
\CodeBefore
\cellcolor{red!15}{2-7,7-2}
\Body
0.991 	&	0.000 	&	0.000 	&	0.000 	&	0.000 	&	0.132 	&	0.000 	&	0.000 	\\
0.000 	&	0.024 	&	0.000 	&	0.000 	&	0.000 	&	0.000 	&	1.000 	&	0.000 	\\
0.000 	&	0.000 	&	0.997 	&	0.000 	&	0.000 	&	0.000 	&	0.000 	&	0.072 	\\
0.000 	&	0.000 	&	0.000 	&	1.000 	&	0.000 	&	0.000 	&	0.000 	&	0.000 	\\
\hline
0.000 	&	0.000 	&	0.000 	&	0.000 	&	1.000 	&	0.000 	&	0.000 	&	0.000 	\\
0.132 	&	0.000 	&	0.000 	&	0.000 	&	0.000 	&	0.991 	&	0.000 	&	0.000 	\\
0.000 	&	1.000 	&	0.000 	&	0.000 	&	0.000 	&	0.000 	&	0.024 	&	0.000 	\\
0.000 	&	0.000 	&	0.072 	&	0.000 	&	0.000 	&	0.000 	&	0.000 	&	0.997 		
\end{bNiceArray},\nonumber
\end{align}
and the optimal parameters are 
\begin{align}
t_1&=682.06,\ t_2=1095.18,\ t_3=407.60;\nonumber\\
\phi_2&=1.83, \ \phi_3=3.11.\nonumber
\end{align}
The norm of the composite pulse, $U_{CP}$, which pumps $\vert g \rangle \vert 2\rangle$ to $\vert e \rangle \vert 3\rangle$ post-optimization, is
\begin{align}
\arrayrulecolor{cyan}
\begin{bNiceArray}{cccc|cccc}[margin]
\CodeBefore
\cellcolor{red!15}{3-8,8-3}
\Body
0.990 	&	0.000 	&	0.000 	&	0.000 	&	0.000 	&	0.145 	&	0.000 	&	0.000 	\\
0.000 	&	0.999 	&	0.000 	&	0.000 	&	0.000 	&	0.000 	&	0.042 	&	0.000 	\\
0.000 	&	0.000 	&	0.074 	&	0.000 	&	0.000 	&	0.000 	&	0.000 	&	0.997 	\\
0.000 	&	0.000 	&	0.000 	&	1.000 	&	0.000 	&	0.000 	&	0.000 	&	0.000 	\\
\hline
0.000 	&	0.000 	&	0.000 	&	0.000 	&	1.000 	&	0.000 	&	0.000 	&	0.000 	\\
0.145 	&	0.000 	&	0.000 	&	0.000 	&	0.000 	&	0.990 	&	0.000 	&	0.000 	\\
0.000 	&	0.042 	&	0.000 	&	0.000 	&	0.000 	&	0.000 	&	0.999 	&	0.000 	\\
0.000 	&	0.000 	&	0.997 	&	0.000 	&	0.000 	&	0.000 	&	0.000 	&	0.074 			
\end{bNiceArray},\nonumber
\end{align}
and the optimal parameters are  
\begin{align}
t_1&=846.07,\  t_2=1088.08,\ t_3=1265.53, \nonumber\\
\phi_2&=4.50,\ \phi_3=5.66. \nonumber
\end{align}

 Utilizing the three composite pulses outlined in Fig.~\ref{fig:fig1}, we are able to measure the population of the Fock states $\vert 0 \rangle$, $\vert 1 \rangle$, and $\vert 2 \rangle$. By restricting our analysis to the weak coupling regime and limiting the Boson space dimensions to four, we achieve high fidelity suitable for phonon number measurements.

\section{Strong coupling composite pulses}\label{sec:strong_cp}
In contrast to the weak coupling regime, strong coupling strength, \textcolor{black}{where $|\Omega|$ and $|\Delta|$ are of the same order of magnitude}, induces a significant AC Stark shift, which affects the optimal laser detuning. Within the strong coupling regime, we set $\Omega_1=\Omega_2=\Omega_3=1$ and $\phi_1=0$. The parameters requiring optimization include $\Delta=\Delta_1=\Delta_2=\Delta_3$, $t_1$, $t_2$, $t_3$, $\phi_2$, and $\phi_3$.

\begin{figure}
\centering
\includegraphics[width=0.65\columnwidth]{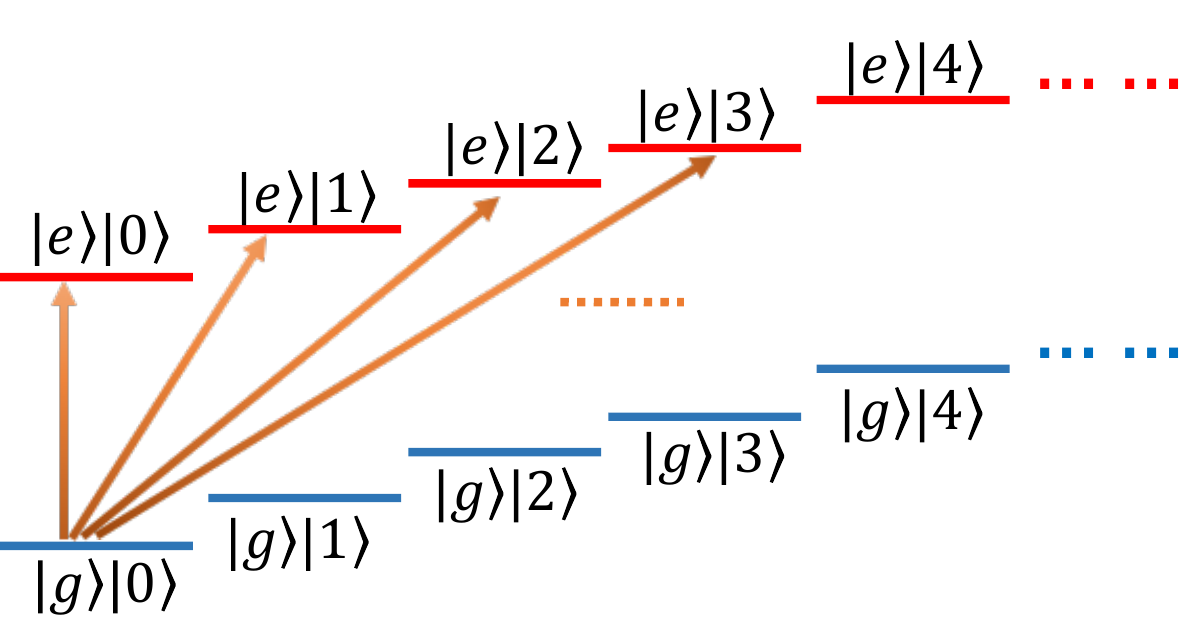}
\caption{The schematic diagram of strong coupling unitary transforming. \textcolor{black}{To measure the population of a Fock state using the electronic shelving technique, it suffices to excite a Fock state $\vert g \rangle \vert n \rangle$ to an excited state $\vert e \rangle$, ignoring the difference in the phonon state.} As depicted, \textcolor{black}{the state $\vert g \rangle \vert 0 \rangle$ is exclusively pumped to an excited state $\vert e \rangle$, except for the difference in the resultant Fock state.} Similarly, states such as $\vert g \rangle \vert 1 \rangle$ and $\vert g \rangle \vert 2 \rangle$ undergo analogous transitions. }
\label{fig:fig3}
\end{figure}

In this case, there is a key difference in the optimization process of Eq. (\ref{eq:loss2}) about the form of the target \textcolor{black}{transition} \textcolor{black}{$U_{target}$}. It is not necessary to limit the composite pulses solely to a SWAP operation. Furthermore, \textcolor{black}{to provide more degrees of freedom for the optimization process,} the unitary operation does not need to satisfy $UU=I$, when focusing on temperature measurement objectives. \textcolor{black}{At the same time, because strong coupling possibly induces higher-order sideband processes, we allow transition forms as shown in Fig.~\ref{fig:fig3} to occur. }To measure the population of a Fock state using the electronic shelving technique, it suffices to excite a Fock state $\vert g \rangle \vert n \rangle$ to an excited state $\vert e \rangle$, \textcolor{black}{ignoring the difference in the phonon state}. Consequently, the target \textcolor{black}{transition} \textcolor{black}{$U_{target}$} in Eq. (\ref{eq:SWAP}) adopts the form in \textcolor{black}{table} (\ref{eq:str_gate}), 
\begin{align}\label{eq:str_gate}
  \textcolor{black}{U_{target}}=\arrayrulecolor{cyan}
  \begin{bNiceArray}{cccc|cw{c}{1cm}cc}[first-col,margin]
  \vert g\rangle \vert 0\rangle & 0 & 0 & 0 & 0 &\Block{4-4}<\Large>{A}  & & &  \\
  \vert g\rangle \vert 1\rangle & 0 & 1 & 0 & 0  & & & & \\
   \vert g\rangle \vert 2\rangle&0&0&1&0&&&&\\
   \vert g\rangle \vert 3\rangle&0&0&0&1&&&&\\
   \hline
   \vert e\rangle \vert 0\rangle &\Block{4-4}<\Large>{B}&&&&\Block{4-4}<\Large>{C}&&&\\
   \vert e\rangle \vert 1\rangle&&&&&&&&\\
   \vert e\rangle \vert 2\rangle&&&&&&&&\\
   \vert e\rangle \vert 3\rangle&&&&&&&&
  \end{bNiceArray},
\end{align}

where blocks A, B, and C are excluded from the loss function calculations, and the comparison between the composite pulses' unitary matrix and the target \textcolor{black}{transition} is confined to the top left quarter. \textcolor{black}{By modifying the top-left corner of the matrix, we can optimize the corresponding transitions. This is done by setting the matrix elements of the Fock states to be detected to 0, while the remaining diagonal elements are set to 1, and the A, B and C blocks are free.}

In a manner analogous to the calculations presented in the preceding section, we give the optimal control parameters and ${\rm norm.}(U_{CP})$ of the unitary matrix for composite pulses. After optimization, the ${\rm norm.}(U_{CP})$ of the composite pulse pumping $\vert g \rangle \vert 0\rangle$ to $\vert e \rangle$  is
\begin{align}
\arrayrulecolor{cyan}
\begin{bNiceArray}{cccc|cccc}[margin]
\CodeBefore
\cellcolor{red!15}{1-5,5-1}
\Body
0.097 	&	0.073 	&	0.007 	&	0.001 	&	0.988 	&	0.090 	&	0.004 	&	0.001 	\\
0.039 	&	0.862 	&	0.058 	&	0.017 	&	0.047 	&	0.414 	&	0.278 	&	0.019 	\\
0.004 	&	0.080 	&	0.911 	&	0.048 	&	0.005 	&	0.073 	&	0.371 	&	0.139 	\\
0.001 	&	0.003 	&	0.049 	&	0.883 	&	0.000 	&	0.007 	&	0.073 	&	0.462 	\\
\hline
0.993 	&	0.026 	&	0.006 	&	0.000 	&	0.100 	&	0.056 	&	0.006 	&	0.001 	\\
0.050 	&	0.429 	&	0.021 	&	0.014 	&	0.103 	&	0.891 	&	0.096 	&	0.011 	\\
0.020 	&	0.246 	&	0.378 	&	0.088 	&	0.002 	&	0.136 	&	0.877 	&	0.013 	\\
0.001 	&	0.014 	&	0.145 	&	0.459 	&	0.000 	&	0.005 	&	0.035 	&	0.876 			
\end{bNiceArray},\nonumber
\end{align}
and the optimal parameters are  
\begin{align}
t_1&=148.17,\ t_2=149.50,\ t_3=140.63;\nonumber\\
\phi_2&=1.12, \ \phi_3=6.38,\ \Delta=0.93.\nonumber
\end{align}
Therefore, \textcolor{black}{the position of the largest value of the matrix element modulus indicates} the main process of this composite pulse is the carrier transition.
The ${\rm norm.}(U_{CP})$ of the composite pulse pumping $\vert g \rangle \vert 1\rangle$ to $\vert e \rangle$ after optimization is
\begin{align}
\arrayrulecolor{cyan}
\begin{bNiceArray}{cccc|cccc}[margin]
\CodeBefore
\cellcolor{red!15}{2-8,8-2}
\Body
0.711 	&	0.021 	&	0.060 	&	0.009 	&	0.012 	&	0.044 	&	0.698 	&	0.010 	\\
0.027 	&	0.139 	&	0.038 	&	0.030 	&	0.011 	&	0.051 	&	0.038 	&	0.987 	\\
0.050 	&	0.077 	&	0.990 	&	0.025 	&	0.029 	&	0.027 	&	0.087 	&	0.037 	\\
0.011 	&	0.055 	&	0.022 	&	0.997 	&	0.001 	&	0.047 	&	0.005 	&	0.024 	\\
\hline
0.037 	&	0.005 	&	0.028 	&	0.001 	&	0.999 	&	0.005 	&	0.024 	&	0.011 	\\
0.041 	&	0.024 	&	0.026 	&	0.046 	&	0.006 	&	0.996 	&	0.027 	&	0.054 	\\
0.698 	&	0.023 	&	0.083 	&	0.009 	&	0.042 	&	0.022 	&	0.708 	&	0.040 	\\
0.004 	&	0.985 	&	0.075 	&	0.051 	&	0.007 	&	0.029 	&	0.028 	&	0.140 			
\end{bNiceArray},\nonumber
\end{align}
and the optimal parameters are 
\begin{align}
t_1&=145.29, \ t_2=149.08,\ t_3=144.11;\nonumber\\
\phi_2&=5.58,\ \phi_3=6.57,\ \Delta=1.73. \nonumber
\end{align}
The primary dynamical process of this composite pulse is the second-order blue-sideband transition. It is evident that the \textcolor{black}{transition fidelity of the composite pulse} is lower compared to the weak coupling scenario; therefore, we propose a modified method in Sec. \ref{sec:modify}. The operator $U_{CP}$, which pumps $\vert g \rangle \vert 2\rangle$ to $\vert e \rangle$ and exhibits improved performance after optimization, has the form
\begin{align}
\arrayrulecolor{cyan}
\begin{bNiceArray}{cccc|cccc}[margin]
\CodeBefore
\cellcolor{red!15}{8-3,7-3,3-8,3-7}
\Body
0.964 	&	0.144 	&	0.010 	&	0.001 	&	0.168 	&	0.146 	&	0.020 	&	0.002 	\\
0.138 	&	0.966 	&	0.032 	&	0.007 	&	0.030 	&	0.192 	&	0.093 	&	0.022 	\\
0.010 	&	0.071 	&	0.046 	&	0.015 	&	0.005 	&	0.047 	&	0.886 	&	0.453 	\\
0.001 	&	0.013 	&	0.059 	&	0.971 	&	0.001 	&	0.006 	&	0.109 	&	0.205 	\\
\hline
0.169 	&	0.003 	&	0.001 	&	0.000 	&	0.985 	&	0.041 	&	0.005 	&	0.001 	\\
0.151 	&	0.190 	&	0.058 	&	0.010 	&	0.033 	&	0.967 	&	0.027 	&	0.021 	\\
0.003 	&	0.066 	&	0.871 	&	0.161 	&	0.001 	&	0.045 	&	0.241 	&	0.388 	\\
0.002 	&	0.029 	&	0.481 	&	0.176 	&	0.002 	&	0.026 	&	0.368 	&	0.775 			
\end{bNiceArray},\nonumber
\end{align}
and the optimal parameters are 
\begin{align}
t_1&=149.89,\ t_2=140.06,\ t_3=144.92; \nonumber\\
\phi_2&=6.87,\ \phi_3=1.51,\ \Delta=0.73.\nonumber
\end{align}
\textcolor{black}{Based on the positions of the largest values in the above matrix, we can identify the main transition processes} of this composite pulse is the carrier transition. 

In the strong coupling regime, the transitions stimulated by composite pulses differ significantly from those in the weak coupling regime. These transitions include carrier transitions, as well as first and second blue-sideband transitions. These results, designed through quantum optimal control, has considered non-resonant excitation processes and all orders of exponential expansion in the creation and annihilation of bosons. Additionally, they account for the AC Stark shift caused by the laser, which cannot be computed using analytical methods.

 \section{Modification and examples}\label{sec:modify}

We can enhance the theoretical \textcolor{black}{transition fidelity of the composite pulse} by incorporating additional pulses; however, this increases computational resource consumption and may reduce experimental robustness. Consequently, a modification in the \textcolor{black}{simulation} approach is necessary. We implement our scheme in a typical scenario where, following the cooling of a thermal state, the phonon state exhibits a super-Poissonian distribution \cite{Walls2008}. In this distribution, the phonon population declines sharply as the Fock state number increases.

After cooling, the ion's internal state is the ground state $\vert g \rangle$. Our scheme is thus applicable for measuring the diagonal elements of the density matrix in the Fock basis \textcolor{black}{when the phonon state is in a thermal distribution}. Implementing a composite pulse on the initial state results in a non-zero phonon population in the excited internal state. We then employ the electronic shelving method to measure this population. However, the obtained value does not solely represent the population of the anticipated excited state $\vert g \rangle \vert n \rangle$, as other levels are also excited due to the imperfection of the composite pulse. This necessitates further modifications to enhance accuracy. Additionally, the effectiveness of our method is constrained to cases with a centralized phonon distribution, such as thermal states, due to the finite number of pulses.

The modification process is described by
\begin{align}\label{eq:modify}
\begin{bNiceMatrix}
a_{00} & a_{01} & \Cdots &a_{0n}\\
a_{10} & a_{11} & \Cdots &a_{1n}\\
\Vdots & \Vdots &\Ddots & \Vdots\\
a_{n0} & a_{n1} & \Cdots &a_{nn}
\end{bNiceMatrix}
\begin{bNiceMatrix}
P_0 \\
P_1 \\
 \Vdots \\
  P_n
\end{bNiceMatrix}
=
\begin{bNiceArray}{c}
M_0\\
M_1 \\
\Vdots \\
 M_n
\end{bNiceArray},
\end{align}
where $P_n$ represents the \textcolor{black}{true} phonon population in the Fock state $\vert n \rangle$. The element $M_n$ on the right side of the equation corresponds to the measurement value obtained via the electronic shelving method. The element $a_{mn}$ of the left coefficient matrix indicates the proportion of the state $\vert g \rangle \vert n \rangle$ being excited to the state $\vert e \rangle$ for any phonon state by the composite pulse $m$. For instance, consider a composite pulse designed primarily to excite $\vert g \rangle \vert 1 \rangle$ to $\vert e \rangle \vert 2 \rangle$. This pulse is represented by the unitary matrix $U$. Consequently, $a_{11}=\sum_{i=1}^{\frac{n}{2}}\vert U_{\frac{n}{2}+i,2} \vert ^2$ and $a_{12}=\sum_{i=1}^{\frac{n}{2}}\vert U_{\frac{n}{2}+i,3} \vert ^2$, where $n$ represents the number of rows in matrix $U$.

\begin{figure}[h!]
 \centering
 \subfloat[]
  {
      \label{fig:subfig1}\includegraphics[width=0.5\textwidth]{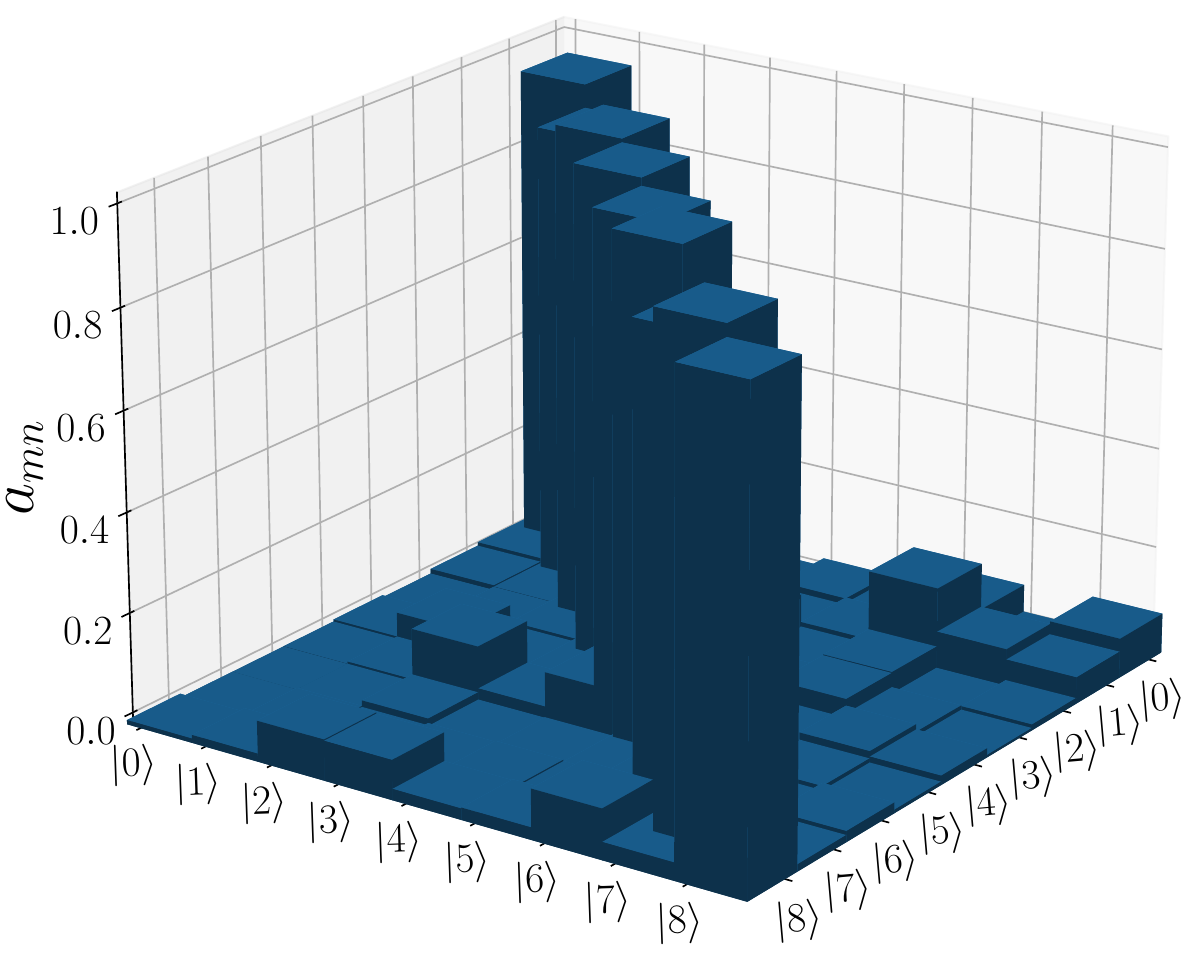}
  }
  \subfloat[]
  {
      \label{fig:subfig2}\includegraphics[width=0.5\textwidth]{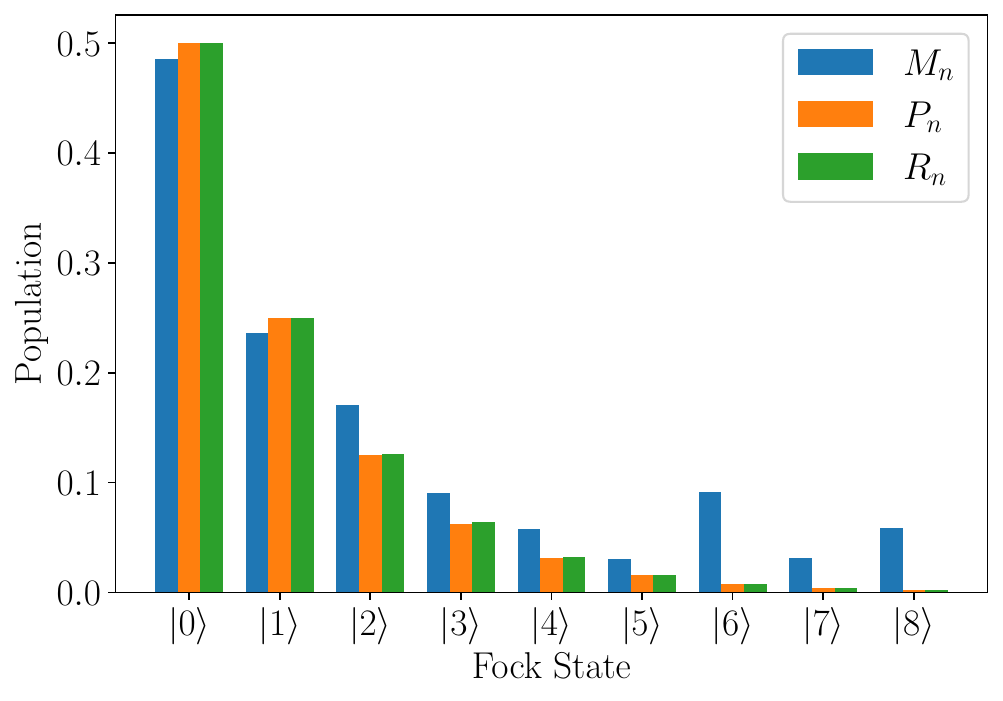}
  } 
  \caption{The bar chart of the Fock state populations of $P_n$, $M_n$ in Eq.(\ref{eq:modify}) and the \textcolor{black}{modified estimate results} $R_n$ in the weak coupling regime. We observe that while the direct measurement results, $M_n$, exhibit discrepancies with the \textcolor{black}{true} population values, $P_n$, the \textcolor{black}{modified estimate results}, $R_n$, closely approximate $P_n$.} 

 \label{fig:fig4} 
 \end{figure}

\begin{figure}[h!]
 \centering
 \subfloat[]
  {
      \label{fig:subfig3}\includegraphics[width=0.5\textwidth]{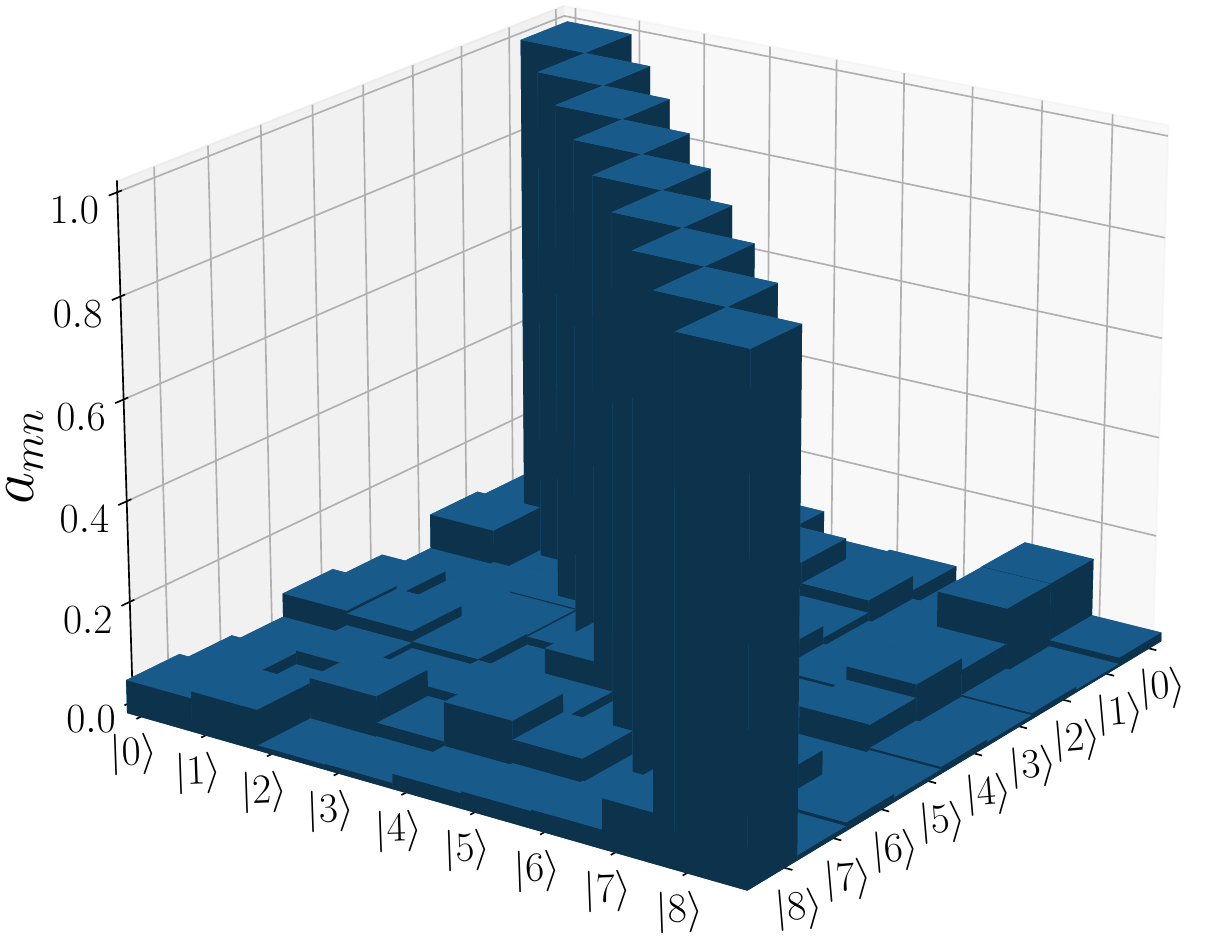}
  }
\subfloat[]
  { 
      \label{fig:subfig4}\includegraphics[width=0.5\textwidth]{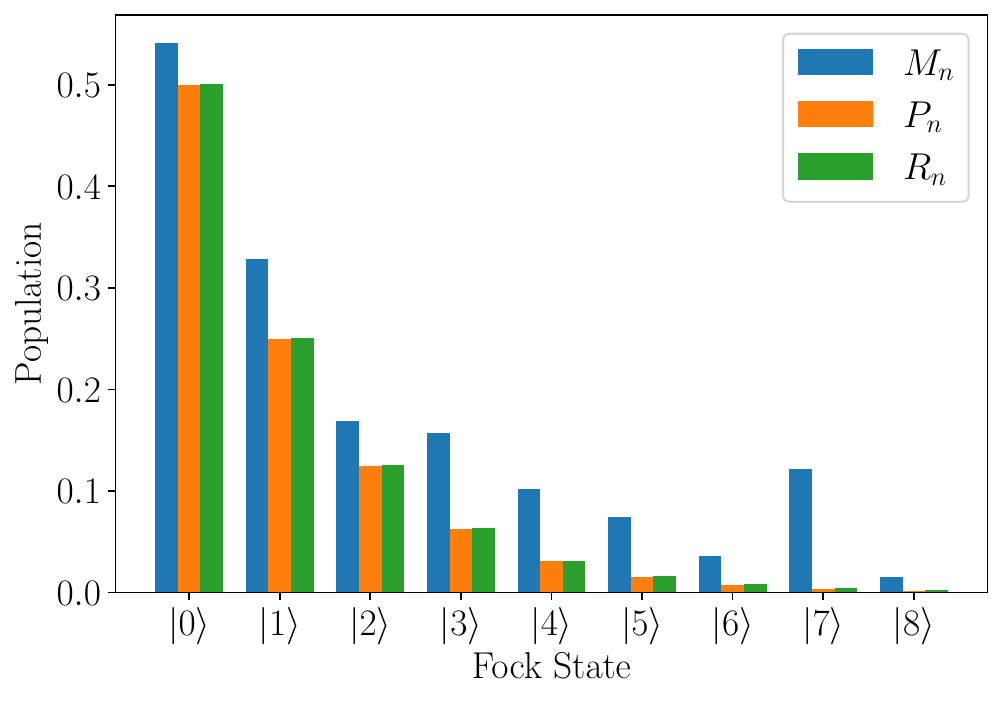}
  }
  \caption{The bar chart of the Fock state populations of $P_n$, $M_n$ in Eq.(\ref{eq:modify}) and the \textcolor{black}{modified estimate results} $R_n$ in the strong coupling regime, \textcolor{black}{where we set $\Omega = 1$, then the optimized value of $\Delta$ is around 1}.}   
  \label{fig:fig5} 
 \end{figure}

Suppose that there exists an ion in an approximate thermal state after cooling, with an average phonon number of 1.0, indicating suboptimal cooling. \textcolor{black}{One can utilize multiple composite pulses to individually read the first few Fock states in order to fit a thermal state.} We limit the Boson space to 10 dimensions and employ a weak coupling laser for composite pulse manipulation. Due to the large size of the 10-dimensional matrix, we present only the coefficient matrix \textcolor{black}{in Fig. \ref{fig:fig4} and Fig. \ref{fig:fig5}}. We also compare the actual thermal distribution with the measurements.

Unlike the scenarios described in Secs.~\ref{sec:SWAP} and \ref{sec:strong_cp}, where the Boson subspace was truncated to four dimensions with five parameters to optimize, we now truncate the Boson subspace to 10 dimensions. Consequently, the original five parameters must now satisfy 10 equations, which is unfeasible. Therefore, we utilize six pulses instead.

To simulate a relatively realistic situation, we computed the distribution for a Boson cutoff of 100, which we consider as the real values without any cutoff, as this cutoff negligibly influences the distribution. We then truncate the Boson subspace to 10 for quantum optimal control optimization of the six-pulse composite pulses, simulating finite dimension truncation. The coefficient matrix is derived from this 10-dimensional truncation. \textcolor{black}{The truncation of the coefficient matrix may cause error, we use $R_n$ in the figures below to represent the modified estimate result as a comparison to the true value $P_n$, where $\bm{R}=\bm{a}^{-1}\cdot\bm{M}$}

Fig.~\ref{fig:fig4} illustrates that the error in direct measurement is tolerable around the ground state of the phonon, and the modification yields results that closely approximate the actual values. Fig.~\ref{fig:fig5} demonstrates a similar effect of our method in the strong coupling regime. It is important to note that when employing six-pulse composite pulses for quantum optimal control processes, the fidelity of each composite pulse is significantly enhanced. Moreover, the primary transitions between $\vert g \rangle$ and $\vert e \rangle$ are predominantly first-order blue-sideband transitions, which differ markedly from the optimal results discussed in Sec.~\ref{sec:strong_cp}. These examples demonstrate that the proposed modification method can enhance the accuracy of phonon population measurements.

\begin{figure}[htbp]
  \includegraphics[width=0.65\columnwidth]{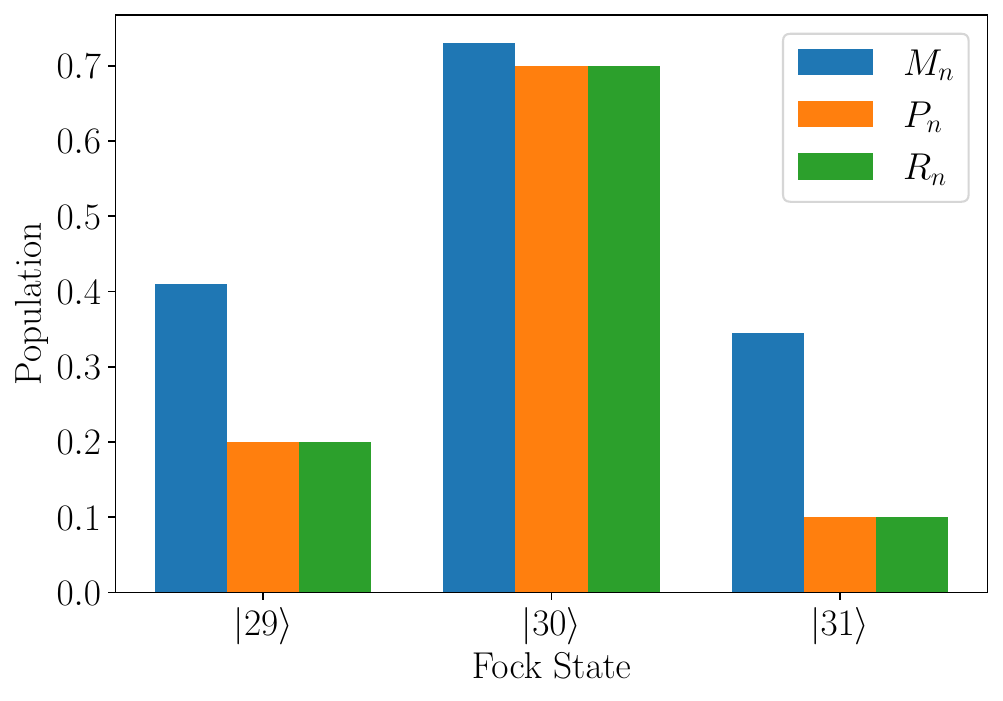}
  \centering
  \caption{The example of high Fock state measurement. The actual population of phonon is $P_{29}=0.3$, $P_{30}=0.4$, $P_{31}=0.3$. After the measuring pulses and modification, the result is very close to the actual distribution.}
  \label{fig:fig6}
\end{figure}

Next, we provide an example to illustrate the advantage of our scheme in measure high Fock state population. In Ref.~\cite{Um:2016aa}, the authors mainly use the physical fact that Fock state has no negative number, so the red-sideband pulse can not pump $\vert g \rangle \vert 0 \rangle$ to excited state. Comparing with the measuring process in Ref.~\cite{Um:2016aa}, our method can directly measure the higher Fock state population \textcolor{black}{instead} of sequentially measuring the population of $\vert g \rangle \vert 0 \rangle$, $\vert g \rangle \vert 1 \rangle$, $\vert g \rangle \vert 2 \rangle$ and so on. Suppose there is a phonon state which populates mainly in $\vert g \rangle \vert 29 \rangle$, $\vert g \rangle \vert 30 \rangle$, $\vert g \rangle \vert 31 \rangle$, and $P_{29}=0.3$, $P_{30}=0.4$, and $P_{31}=0.3$. Fig.~\ref{fig:fig6} gives the result of Fock state detection using optimized composite pulses where the population we get after modification is consistent with its real values.

\section{Conclusion}\label{sec:summary}
 To summarize, we have proposed a method for measuring phonon numbers using composite pulses enhanced by quantum optimal control. We optimized the weak and strong coupling regimes separately through quantum optimal control, discovering that optimal parameters lead to distinct physical processes. In the weak coupling regime, the blue-sideband process is optimal, while in the strong coupling regime, either the carrier or the second-order blue-sideband transformation can be utilized. We provide a concrete example of how our proposed method can be used to measure the phonon number in a thermal state phonon, and we discuss modifications to enhance the measurement accuracy. Additionally, we highlight the advantage of our scheme, which allows for the direct measurement of phonon populations concentrated in a high Fock state, eliminating the need for adiabatic evolution of a high Fock state to $\vert g \rangle \vert 0 \rangle$ step by step.

Further work is required in several areas. First, our method can measure Fock state populations, and by incorporating some unitary transformations, it can be adapted for phonon state tomography. Second, the robustness of the composite pulses can be further enhanced. \textcolor{black}{Third, if the distribution of the phonon is too broad, and the evolution time of the composite pulses can be long, then the ions may be heated during the measuring process, and how to construct fast pulses is a problem that needs to be addressed.} Fourth, we have considered only single-ion scenarios. In systems with $N$ ions, each direction has $N$ motional modes, and \textcolor{black}{the composite pulse used to detect a specific phonon mode may be influenced by other modes}, particularly in the strong coupling regime. 



\begin{acknowledgments}
  This work is supported by National Natural Science Foundation of China under Grants No. 12074433. 
\end{acknowledgments}

\bibliographystyle{apsrev4-1}
\bibliography{refs}
 
\end{document}